\documentclass[journal,twoside]{IEEEtran}
\usepackage{cite}
\usepackage{amsmath,amssymb,amsfonts}
\usepackage{algorithmic}
\usepackage{graphicx}
\usepackage{textcomp}
\usepackage{xcolor}
\usepackage{subcaption}
\usepackage{multirow}
\usepackage{hyperref}

\def\BibTeX{{\rm B\kern-.05em{\sc i\kern-.025em b}\kern-.08em
    T\kern-.1667em\lower.7ex\hbox{E}\kern-.125emX}}
\begin{document}
\title{Uncertainty-Aware Correction and Language-Guided Refinement for Brain Tumor Segmentation in MRI}
\author{Danish Ali, Ajmal Mian, Naveed Akhtar, and Ghulam Mubashar Hassan
\thanks{
This research is supported by the Australian Government
Research Training Scholarship. The authors also gratefully acknowledge the organizers of the BraTS2023 Challenge for providing the dataset used in this research. Dr. Naveed Akhtar is a recipient of the ARC Discovery Early Career Researcher Award (project \#DE230101058), funded by the Australian Government. }
\thanks{Danish Ali, Ajmal Mian, and Ghulam Mubashar Hassan are with The University of Western Australia, Perth, WA 6009, Australia (e-mail: danish.ali@research.uwa.edu.au; ajmal.mian@uwa.edu.au; ghulam.hassan@uwa.edu.au).}
\thanks{Naveed Akhtar is with The University of Melbourne, Melbourne, Parkville VIC 3010, Australia (e-mail: naveed.akhtar1@unimelb.edu.au).}
\thanks{Corresponding Author: Danish Ali.}
}

\maketitle
\begin{abstract}
Accurate brain tumor segmentation in magnetic resonance images (MRI) is essential for
clinical assessment and treatment planning. Existing methods have largely focused on improving the initial segmentation
accuracy, with limited emphasis on identifying
and correcting errors in the segmentation predictions. We propose an
uncertainty-aware brain tumor segmentation network
that integrates automatic error correction and
language-guided refinement within a four stage design. The first stage generates an initial segmentation
using a prompt-free 3D segmentation backbone,
while the second stage performs targeted refinement
through language prompts.
We propose a correction token generation module (stage 3)
that automatically generates correction prompts from the initial segmentation predictions 
of the MRI, to facilitate refinement 
at inference. Furthermore, a Windowed Masked Correction Attention
(WMCA) module (stage 4) enables interactive refinement through
uncertainty-aware region selection and targeted
class transitions guidance. Extensive evaluations across four benchmarks,
including BraTS 2020 Glioma, BraTS 2023 Glioma, and two external cohorts (BraTS 2023 Meningioma and BraTS 2023 Pediatric), demonstrate that our method outperforms
state-of-the-art approaches in several comparisons,
achieving higher Dice scores and lower HD95 distances. Paired Wilcoxon signed-rank tests further
demonstrate the statistical significance of
the observed performance improvements.

\end{abstract}
\begin{IEEEkeywords}
Brain Tumor Segmentation, Text-Guided Correction, Interactive Segmentation
\end{IEEEkeywords}
\section{Introduction}
\label{sec:introduction}

Brain tumors are life-threatening conditions that require
accurate diagnosis and effective treatment planning~\cite{menze2014multimodal}. Multi-parametric magnetic resonance imaging (mpMRI)
provides complementary information through four sequences:
T1, contrast-enhanced T1 (T1ce),
T2, and fluid-attenuated inversion recovery
(FLAIR)~\cite{zhou2021latent}. Precise segmentation of tumor sub-regions: necrotic core, edema, and enhancing tumor, 
from these sequences is essential for tumor assessment,
surgical planning, and longitudinal monitoring~\cite{lu2026pgr}. However, manual delineation is labor-intensive and
time-consuming, particularly for volumetric scans with
complex tumor sub-regions~\cite{sharma2019automated, tao2026mhmamba}. To address this challenge,
deep learning based methods have significantly advanced
automated segmentation, enabling efficient delineation
of clinically relevant tumor regions~\cite{havaei2017brain, she2023eoformer, ali2025attention, zhang2025incomplete}. 

Architectures have evolved from convolutional neural
networks (CNNs) to Transformers, state-space models,
and hybrid designs (Sec.~\ref{sec:related_work}),
yet they primarily focus on improving the initial
segmentation accuracy, with limited emphasis on
identifying and correcting residual prediction errors.
In clinical applications, inaccurate delineation can
compromise the assessment of tumor extent and
surrounding anatomy, potentially affecting critical
treatment decisions, and experts may need to inspect
and correct imperfect automated segmentations,
introducing additional manual effort~\cite{hoebel2024not}. This calls for methods that can
automatically identify potential errors and
perform targeted refinement, with {\em optional}
human intervention to further improve segmentation
accuracy by specifying the desired target classes
for identified uncertain regions.

Addressing the clinical requirement, we propose a first-of-its-kind  uncertainty-aware brain
tumor segmentation network that integrates automatic
error correction and language-guided refinement. A prompt-free 3D
segmentation backbone first generates an initial
segmentation, which is then refined using targeted
correction prompts. Our main contributions are:
\begin{itemize}
    \item A language-guided correction network that
    refines initial segmentation predictions using
    spatially structured instructions specifying
    error locations, directed class transitions,
    and correction extents.
    \item A correction token generation module that
learns to generate correction prompts
from MRI, initial segmentation predictions,
and their associated uncertainty, enabling
targeted refinement without ground-truth
access at inference.

    \item A Windowed Masked Correction Attention 
    module that enables interactive refinement through
    uncertainty-aware region selection, where  users
    can directly  specify the desired target classes for the
    identified uncertain regions, enabling minimal effort targeted intervention.
\end{itemize}

We conduct extensive evaluations on two BraTS glioma benchmarks (BraTS 2020 and BraTS 2023)~\cite{bakas2018identifying,baid2021rsna} and two external BraTS cohorts (Meningioma and Pediatric Brain Tumor)~\cite{kazerooni2024brain,labella2023asnr} to assess segmentation performance and cross-tumor generalization. The experimental results demonstrate that our method outperforms state-of-the-art approaches on several segmentation metrics, achieving
Dice gains of 0.29-1.36 percentage points and HD95
reductions of 0.05-0.93 mm, with statistical
significance confirmed by paired Wilcoxon
signed-rank tests.

\section{Related Work}
\label{sec:related_work}

\noindent\textbf{CNN-based segmentation.}
Early automated brain tumor segmentation relied on
2D CNNs~\cite{ottom2022znet, metlek2023resunet+},
with U-Net~\cite{ronneberger2015u} as a foundational
architecture that preserves fine-grained spatial
information through skip connections.
3D U-Net~\cite{cciccek20163d} extended convolutional
operations to volumetric images to model
three-dimensional spatial dependencies, and subsequent
3D CNN-based methods leveraged volumetric features to
improve tumor delineation~\cite{zhou2020afpnet, anand2020brain}.
However, their local receptive fields limit the
modeling of long-range spatial
dependencies~\cite{chen2021transunet}.

\noindent\textbf{Transformer-based segmentation.}
Transformer-based methods leverage self-attention to
model global contextual information.
TransBTS~\cite{wang2021transbts} integrates 3D CNNs
with Transformers to capture both local and global
features, while UNETR~\cite{hatamizadeh2022unetr}
employs a Transformer encoder to learn global
representations at multiple spatial resolutions.
mmFormer~\cite{zhang2022mmformer} uses intra-modal and
inter-modal Transformers to model long-range
dependencies within and across MRI modalities.
To reduce the burden of global self-attention,
Swin UNETR~\cite{hatamizadeh2021swin} employs
hierarchical shifted-window attention, although its computational cost remains quadratic in the number of tokens within
each window~\cite{liu2021swin}.

\noindent\textbf{State-space and hybrid models.}
State-space models, particularly Mamba~\cite{gu2023mamba},
model long-range dependencies with linear complexity
in sequence length. SegMamba~\cite{xing2025segmamba}
introduces tri-oriented spatial Mamba blocks capturing
forward, reverse, and inter-slice feature interactions.
SegMamba-V2~\cite{xing2025segmambav2} extends them
to tri-oriented orthogonal blocks across axial,
sagittal, and coronal planes, at the cost of
substantial overhead from repeated feature scanning.
DRBD-Mamba~\cite{ali2026drbd} reduces this drawback
with a dual-resolution bidirectional design that uses
space-filling curves to preserve spatial locality, while
MHMamba~\cite{tao2026mhmamba} introduces parallel Mamba
heads with channel-space calibration and adaptive
skip-connection fusion. Hybrid architectures combine
complementary modeling mechanisms:
IM-Fuse~\cite{pipoli2025fuse} integrates
Transformer-based contextual modeling with Mamba-based
multimodal feature fusion, whereas
UniME~\cite{song2026uni} combines a pretrained unified
Transformer encoder with modality-specific CNN encoders.

\noindent\textbf{Error identification and correction.}
The above methods offer limited mechanisms for
automatically identifying and correcting prediction
errors, leaving residual mistakes to manual
expert correction~\cite{hoebel2024not}. Our work
instead targets the step after initial segmentation:
estimating where the prediction is likely wrong and
correcting it through structured prompts that are
generated automatically or provided interactively.

\begin{figure*}[t]
  \centering
  \includegraphics[width=\textwidth]{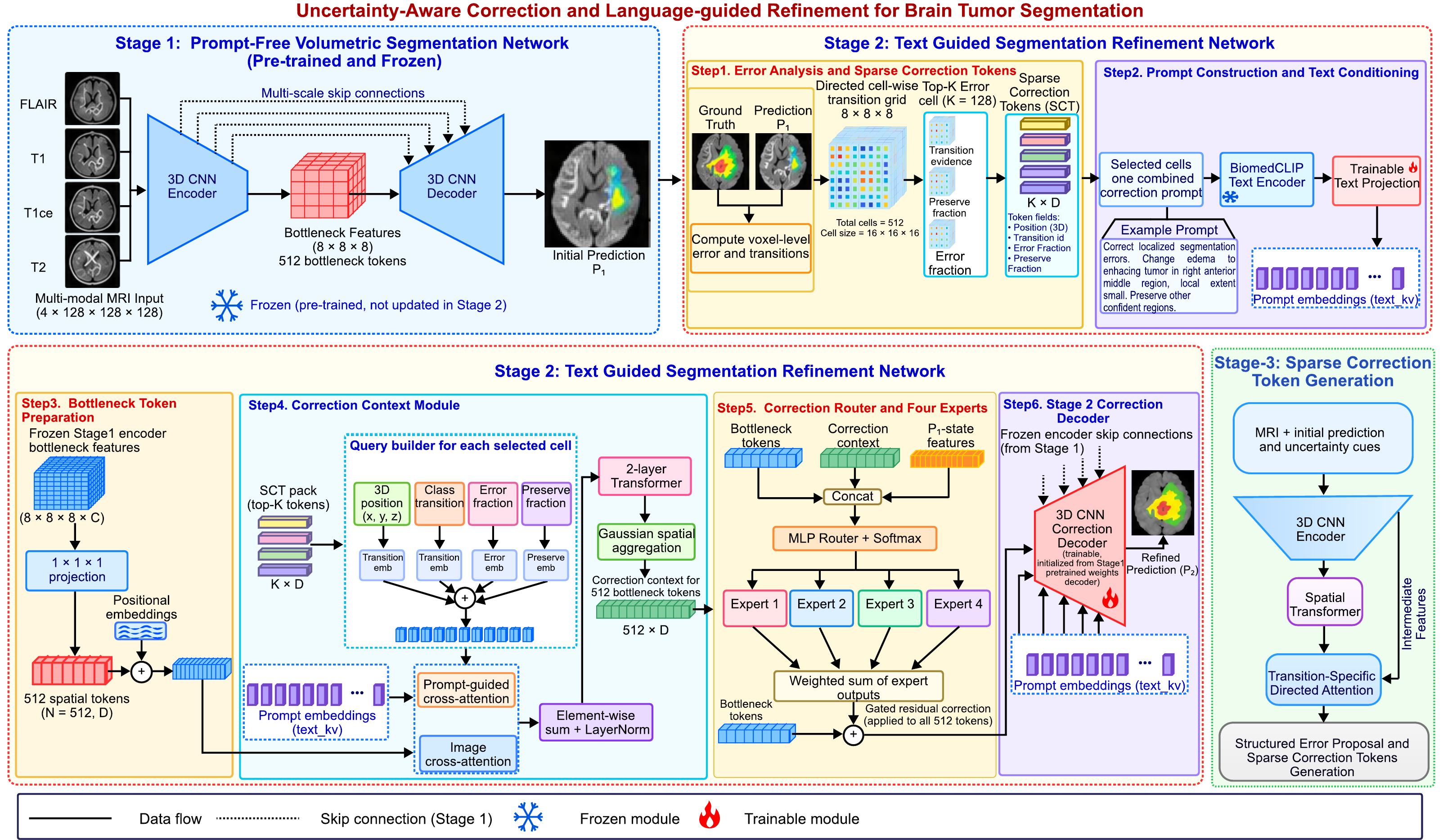}
  \caption{Overview of the proposed architecture for brain tumor segmentation and fully automated uncertainty-aware language-guided refinement, illustrating the three main stages of the segmentation and correction pipeline. Stage 1 generates an initial prediction, while Stage 2 learns language-guided refinement using correction prompts derived from error maps. Stage 3 automatically generates correction prompts from the initial prediction and uncertainty, which guide the trained refinement network to produce the final segmentation.}
  \label{fig:arch}
\end{figure*}
\section{Methodology}
\label{sec:methodology}
Our uncertainty-aware segmentation network learns to identify potential segmentation errors and selectively refine predictions through sparse, spatially localized correction prompts. As illustrated in Fig.~\ref{fig:arch}, it comprises three stages: a 3D U-Net-based network for initial segmentation, a text-guided refinement stage, and an automatic correction prompt generation stage. To bridge the gap toward clinical practice, we propose Stage 4 which is a simulated real-world environment featuring a human-in-the-loop (potentially a clinician-in-the-loop) framework, as presented
in Fig.~\ref{fig:clinctrl}.  

In Stage 1, the segmentation network relies exclusively on multi-modal MRI features to generate an initial prediction $P_1$, which serves as the foundation for the subsequent correction stages. 
In Stage 2, we introduce a text-guided correction network that learns to refine $P_1$ using ground truth-derived correction instructions encoding the locations, class transitions and spatial extents of segmentation errors. 

At inference, Stage 2 receives automatically generated
correction prompts instead of ground-truth-derived
instructions. To this end, we introduce an
uncertainty-aware correction token generation module
in Stage 3 that learns to identify potential segmentation
inaccuracies in $P_1$ and generate targeted correction
prompts. These prompts guide the pretrained Stage 2
correction network to produce a refined segmentation
$P_2$ without requiring ground truth.

In addition to automatic correction, we propose a clinician-in-the-loop refinement pathway (Stage 4), illustrated in Fig.~\ref{fig:clinctrl}, that simulates user intervention by providing target-class correction instructions for regions automatically identified as uncertain in the initial segmentation. This enables targeted refinement at inference time and highlights its potential for clinical applications.
Automatically generated and simulated user-provided
instructions are processed through their respective
correction mechanisms to produce the refined
segmentation $P_2$. The architectural design and
key components of each stage are described in
the following subsections.

\subsection{Stage1: Prompt-Free Volumetric Segmentation}
Given a multi-modal brain MRI volume $\mathbf{X}\in\mathbb{R}^{C\times H\times W\times D}$, where $C$ denotes the number of imaging modalities and $(H,W,D)$ represent the spatial dimensions, we propose a prompt-free 3D U-Net-based architecture for initial segmentation. The network comprises a hierarchical convolutional encoder, a Transformer-based bottleneck, and a convolutional decoder with skip connections. The encoder extracts multi-scale volumetric representations, while the Transformer bottleneck captures long-range dependencies. 

To further enhance contextual feature aggregation, we incorporate window-based attention into the first two decoder stages, enabling localized feature interactions with reduced computational complexity compared to global self-attention. In both the Transformer bottleneck and window-based attention modules, queries, keys, and values are computed exclusively from MRI-derived features, without incorporating external textual prompts. The decoder progressively restores spatial resolution by integrating encoder features through skip connections. Finally, a voxel-wise classification layer produces the initial segmentation prediction $P_1$. The resulting prediction and intermediate feature representations serve as the foundation for the subsequent stages.

\subsection{Stage 2: Text Guided Segmentation Refinement}
\label{sec:prompt_refinement}

Given the initial segmentation prediction $P_1$, we propose a text-guided refinement mechanism that learns to perform targeted corrections using spatially localized instructions. As illustrated in Fig.~\ref{fig:arch}, this refinement stage begins with error analysis, followed by prompt construction, text conditioning, and bottleneck token preparation. The resulting representations are integrated through a correction context module, followed by adaptive correction weighting, and correction decoding. The pretrained segmentation encoder and decoder from Stage 1 remain frozen, while the correction modules and a dedicated refinement decoder are trained using ground-truth-derived correction instructions.\\
\noindent\textbf{Directed Error Analysis and Sparse Correction Tokens.}
Given the initial prediction $P_1$ and ground truth $Y$,
we construct directed error maps to encode voxel-wise
class discrepancies. Let $\hat{Y}_1(v)=\arg\max_c P_1^c(v)$
denote the predicted class at voxel $v$. The directed error map for a class transition $a\rightarrow b$ is formulated as
\begin{equation}
E_{a\rightarrow b}(v)=
\mathbb{I}[\hat{Y}_1(v)=a \land Y(v)=b],
\quad a\neq b,
\label{eq:directed_error}
\end{equation}
where $\mathbb{I}[\cdot]$ denotes a binary indicator function identifying voxel-wise misclassifications for each directed class transition, and $a,b\in{0,1,2,3}$ correspond to background, necrotic core, edema, and enhancing tumor, respectively. Since each class can make transition to any of the other three classes, this yields $4\times3=12$ possible directed transitions. The directed error maps $E_{a\rightarrow b}$ are aggregated
into an $8^3$ spatial grid, producing 512 candidate
cells for a $128^3$ input patch. Each cell is
characterized by a transition histogram
$\mathbf{h}_j\in\mathbb{R}^{12}$ capturing the proportion
of voxels associated with each directed class transition.
The aggregated transition proportions define the local
error fraction $e_j$, while $r_j=1-e_j$ quantifies the
proportion of correctly classified voxels within the cell.
The top-$K$ cells (
$K=128$, empirically selected as discussed in sec \ref{sec:experiments}) with the highest error
fractions are selected and represented as sparse
correction tokens:
\begin{equation}
\mathcal{C}_k=
(\mathbf{p}_k,t_k,e_k,r_k,\mathbf{h}_k,m_k),
\label{eq:correction_pack}
\end{equation}
where $\mathbf{p}_k$ denotes normalized spatial
coordinates, $t_k$ the dominant directed transition,
and $m_k\in\{0,1\}$ indicates whether the selected
cell satisfies the minimum error threshold for correction.
The structured attributes are subsequently encoded
into learnable correction representations.
\\
\noindent\textbf{Correction Prompt Construction.}
The sparse correction tokens are translated into natural-language
instructions describing the intended class transitions, approximate
spatial locations, and error extents. These instructions are encoded using a frozen pretrained BiomedCLIP text encoder, followed by a
learnable linear projection:
\begin{equation}
T=W_t\mathcal{E}_{\mathrm{text}}(\mathcal{P}),
\label{eq:text_embedding}
\end{equation}
where $\mathcal{P}$ denotes the textual prompts,
$\mathcal{E}_{\mathrm{text}}$ represents the text encoder, $W_t$ is the learnable projection, and 
($T$) represents the resulting sequence of projected text embeddings.
\\
\noindent\textbf{Correction Context Module.}
Given the selected correction tokens
$\{\mathcal{C}_k\}_{k=1}^{K}$ and textual embeddings
($T$), we construct a correction
context to integrate spatial correction instructions
with anatomical and semantic evidence. The bottleneck
features from the frozen Stage 1 encoder are projected
through a $1\times1\times1$ convolution, flattened, and
augmented with learnable positional embeddings, yielding
${F}\in\mathbb{R}^{N\times d}$ with $N=512$.

A learnable query builder transforms the spatial coordinates, directed class transitions, error extents, and preservation fractions of the selected correction tokens into correction queries ${Q}$. These queries retrieve complementary anatomical and semantic information through parallel cross-attention with the projected MRI bottleneck features ${F}'$ and textual embeddings ${T}'$, respectively. The resulting attention features are fused through residual aggregation and layer normalization, followed by a
Transformer for contextual refinement:
\begin{equation}
H_{f}=\operatorname{LN}\left(
{Q}
+\operatorname{MHA}({Q}, {F}',{F}')
+\operatorname{MHA}({Q},{T}',{T}')
\right),
\label{eq:correction_context}
\end{equation}
where $\operatorname{MHA}$ denotes multi-head cross-attention,
and $H_f$ represents the fused correction features
corresponding to the $K=128$ selected correction tokens.
To propagate this sparse correction context across the
$N=512$ MRI bottleneck tokens, we employ error-weighted
Gaussian spatial aggregation, which determines the
contribution of each correction token to each bottleneck
location based on spatial proximity, local error extent,
and token validity:
\begin{equation}
\begin{aligned}
w_{k,j}
&=m_k e_k
\exp\left(
-\frac{\|p_k-g_j\|_2^2}{2\sigma^2}
\right),\\
C_j
&=\frac{\sum_{k=1}^{K}w_{k,j}H_{f,k}}
{\max\left(\sum_{k=1}^{K}w_{k,j},\epsilon\right)},
\end{aligned}
\label{eq:spatial_correction}
\end{equation}
where $p_k$ and $g_j$ denote the normalized coordinates
of correction token $k$ and bottleneck token $j$,
respectively, and $\sigma$ controls the spatial
influence of each correction. The resulting
$C\in\mathbb{R}^{N\times d_h}$ provides a
spatially aligned correction context for adaptive
bottleneck refinement.\\
\noindent\textbf{Adaptive Bottleneck Refinement.}
Given the spatially aligned correction context $C$,
we adaptively refine the MRI bottleneck representations.
For each bottleneck token $j$, the image features $f_j$,
initial segmentation state $s_j$, and correction
context $C_j$ are combined to predict adaptive
weights for four parallel learnable feature
transformations (experts):
\begin{equation}
\pi_j=\operatorname{Softmax}\left(
\operatorname{MLP}([f_j;s_j;C_j])
\right).
\label{eq:action_weights}
\end{equation}
The transformation outputs are adaptively weighted
to construct a residual feature update, with an
additional gating factor controlling its magnitude:
\begin{equation}
f_j^{ref}=f_j+\tanh(\alpha)
(1-\pi_{j,4})
\sum_{a=1}^{4}\pi_{j,a}\psi_a(f_j),
\label{eq:adaptive_refinement}
\end{equation}
where $\psi_a$ denotes the $a$-th learnable feature
transformation (expert), $\pi_{j,a}$ its predicted weight,
and $\alpha$ a learnable residual scaling parameter.
The refined bottleneck features are subsequently
passed to the correction decoder to produce
the refined segmentation prediction $P_2$.\\
\noindent\textbf{Correction-Aware Decoding.}
The correction decoder is initialized with the pretrained
Stage 1 decoder weights and progressively restores spatial
resolution using multi-scale features from the frozen encoder.
Unlike the Stage 1 decoder, it incorporates window-based
cross-attention and feature-wise linear modulation (FiLM)
to condition decoder features on the correction context. Specifically, window-based cross-attention enables localized
interactions with textual embeddings, while FiLM uses the global
text representation to scale and shift decoder features. The resulting Stage 2 segmentation is obtained as:
\begin{equation}
P_2=\mathcal{D}_{corr}
(\widetilde{F}_b,\mathcal{S},T,t_g),
\label{eq:refined_segmentation}
\end{equation}
where $\mathcal{D}_{corr}$ denotes the correction decoder,
$\widetilde{F}_b$ represents the refined bottleneck features,
$\mathcal{S}$ denotes the frozen encoder skip features,
and $T$ and $t_g$ represent the token-level and
global text embeddings, respectively.

\begin{figure*}[t]
  \centering
\includegraphics[width=0.9\textwidth]{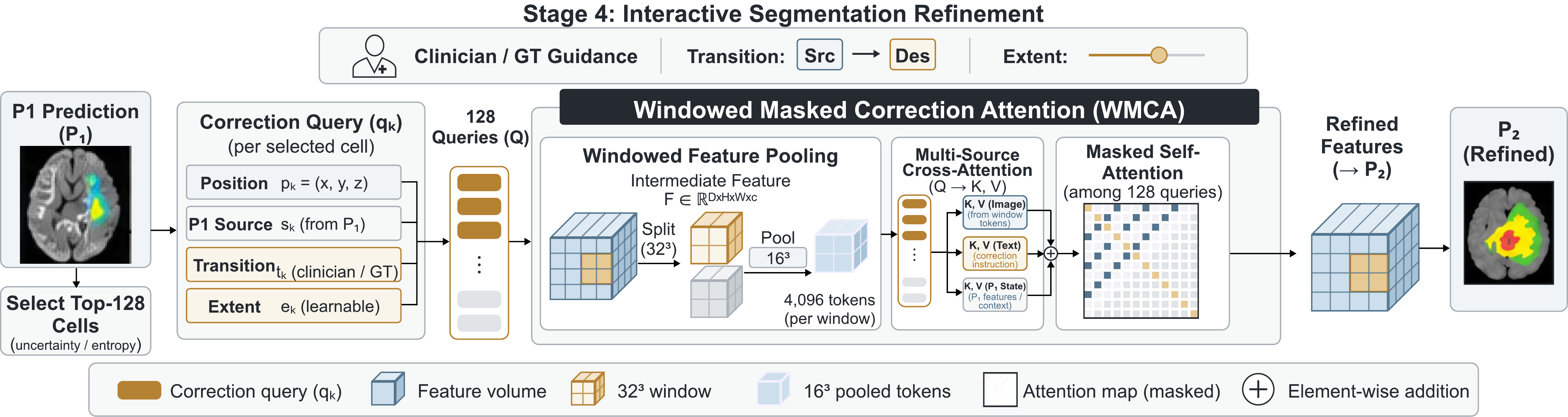}
  \caption{Overview of the proposed interactive segmentation
refinement module (Stage 4), extending the automatic
segmentation refinement framework in
Fig.~\ref{fig:arch} with an optional
clinician-guided correction pathway.
Uncertain cells are selected from the initial
prediction $P_1$ (Stage 1), and simulated clinician
guidance specifies their target classes for
segmentation refinement through Windowed Masked
Correction Attention (WMCA).}
  \label{fig:clinctrl}
\end{figure*}
\subsection{Stage 3: Sparse Correction Token Generation}
Stage 2 enables prompt-guided segmentation
refinement but relies on ground-truth derived
instructions that are unavailable during inference.
To overcome this limitation, we propose a Sparse
Correction Token Generation module that takes multi-modal MRI features and the initial segmentation $P_1$ as input. It employs
Transition-Specific Directed Attention to identify
potential misclassifications between predicted ($P_1$) tumor classes and generate
structured correction tokens for segmentation refinement. The key components of this module are explained below.
\\
\noindent\textbf{Uncertainty-Aware Feature Encoding.}
The four MRI modalities, initial class probabilities ($P_1$), and uncertainty cues comprising prediction
entropy, class-probability margin, and tumor probability
are processed by a 3D CNN encoder. The coarse features
are aggregated into an $8^3$ grid and contextualized
using a spatial Transformer, yielding 512 spatial tokens.
Intermediate CNN features are partitioned into the same
512 spatial cells, each containing 64 local evidence
tokens for subsequent directed attention.
\\
\noindent\textbf{Transition-Specific Directed Attention.}
To explicitly distinguish error directions, we construct
12 transition-specific queries per spatial cell using
its contextualized representation and learnable source,
destination, and direction embeddings. Each query
cross-attends to the corresponding 64 local CNN features,
with attention biased toward regions assigned to its
source class by $P_1$:
\begin{equation}
H_{j,t}=
\operatorname{Softmax}\left(
\frac{Q_{j,t}K_j^\top}{\sqrt{d}}
+\lambda_s\log(O_{j,t}+\epsilon)
\right)V_j,
\label{eq:directed_attention}
\end{equation}
where $H_{j,t}$ denotes the attended feature
representation for transition $t$ in spatial cell
$j$, $Q_{j,t}$ is the corresponding transition-specific
query, and $K_j$ and $V_j$ are the keys and values
projected from local CNN features. $O_{j,t}$ denotes
the source-class occupancy of the local positions
derived from $P_1$, while $\lambda_s$ controls the
strength of the attention bias. The attended
features are used to predict change-versus-preserve
scores, from which source-specific error rates and
destination class probabilities are estimated. Based on the estimated class-transition probabilities
and correction confidence, the top-$K$ ($K=128$)
spatial cells with the highest correction scores
are selected. These cells identify spatial locations and predicted
class transitions for targeted correction, which
are converted into textual instructions using
the same prompt templates as Stage 2. These generated prompts
are then passed to the pretrained correction mechanism
from Stage 2 to refine $P_1$, without
requiring ground-truth annotations at inference.

\subsection{Stage 4: Interactive Segmentation Refinement}\label{interactive-sec}
Beyond the automatic correction framework
illustrated in Fig.~\ref{fig:arch}, we propose
an additional interactive refinement pathway
that enables clinicians to refine the initial
segmentation prediction $P_1$ from Stage 1. To achieve this, we propose Windowed Masked Correction Attention
(WMCA) to identify uncertain regions and guide
targeted refinement. The architecture of the proposed interactive refinement
module is illustrated in Fig.~\ref{fig:clinctrl},
where the intermediate decoder feature map corresponding
to the $128^3$ input patch is partitioned into $16^3$
spatial cells, each covering $8^3$ voxels. The top-$K$ ($K=128$) cells are
selected by their mean voxel-wise prediction entropy
from $P_1$. For each selected cell, a correction
query is constructed by combining learnable
embeddings of its spatial location, source class,
destination class, and correction extent. The spatial location is determined by the cell
coordinates, while the source class is assigned
from $P_1$ using a foreground-priority criterion
based on class occupancy.

Selected cells are assigned to non-overlapping
$32^3$ windows, retaining only those containing
valid correction queries. Within each active window,
multi-source cross-attention integrates intermediate
decoder features, $P_1$ prediction-state features,
and textual embeddings obtained from a frozen BiomedCLIP. Validity-masked self-attention then models
interactions among correction queries belonging
to the same window, while masking queries
associated with other windows:

\begin{equation}
H^{(w)}=\operatorname{Softmax}\left(
\frac{Q^{(w)}K^{(w)\top}}{\sqrt{d}}+M^{(w)}
\right)V^{(w)},
\label{eq:wmca}
\end{equation}
where $Q^{(w)}$, $K^{(w)}$, and $V^{(w)}$
denote the attention projections of correction
representations in window $w$. The mask $M^{(w)}_{ij}$
is set to $0$ for valid key positions and
$-\infty$ otherwise. Gaussian
spatial weighting then propagates correction
information to pooled decoder features, which
undergo refinement to produce
$P_2$ through the segmentation head.

\begin{table*}[t]
\centering
\caption{Quantitative comparison with state-of-the-art methods
across four BraTS benchmarks. Dice (\%, $\uparrow$) and
HD95 (mm, $\downarrow$) are reported for WT, TC, and ET.
Bold and underlined values indicate the best and
second-best performance, respectively.
$^{\ast}$ indicates a statistically significant improvement
over the paired method (two-sided Wilcoxon signed-rank
test, Holm-adjusted $p<0.05$).}
\label{tab:tab1}

\begingroup
\scriptsize
\setlength{\tabcolsep}{1.5pt}
\renewcommand{\arraystretch}{1.0}

\resizebox{\textwidth}{!}{%
\begin{tabular}{@{}lcccccc|cccccc|cccccc|cccccc@{}}

\hline


\rule[-3pt]{0pt}{12pt}
& \multicolumn{6}{c|}{\textbf{BraTS 2020 Glioma}}
& \multicolumn{6}{c|}{\textbf{BraTS 2023 Glioma}}
& \multicolumn{6}{c|}{\textbf{BraTS 2023 Meningioma}}
& \multicolumn{6}{c}{\textbf{BraTS 2023 Pediatric}} \\

\hline


\rule[-3pt]{0pt}{12pt}
& \multicolumn{3}{c}{Dice (\%) $\uparrow$}
& \multicolumn{3}{c|}{HD95 (mm) $\downarrow$}
& \multicolumn{3}{c}{Dice (\%) $\uparrow$}
& \multicolumn{3}{c|}{HD95 (mm)$\downarrow$}
& \multicolumn{3}{c}{Dice (\%) $\uparrow$}
& \multicolumn{3}{c|}{HD95 (mm) $\downarrow$}
& \multicolumn{3}{c}{Dice (\%) $\uparrow$}
& \multicolumn{3}{c}{HD95 (mm) $\downarrow$} \\


\rule[-3pt]{0pt}{12pt}
Method
& WT & TC & ET & WT & TC & ET
& WT & TC & ET & WT & TC & ET
& WT & TC & ET & WT & TC & ET
& WT & TC & ET & WT & TC & ET \\

\hline


DRBD-Mamba~\cite{ali2026drbd}
& 90.83 & 82.64 & 81.58 & 5.34 & 5.96 & 2.65
& 92.79 & 89.96 & 87.50 & 4.14 & 3.77 & 2.00
& 86.54 & 88.54 & 88.73 & 9.63 & 5.60 & 4.97
& 84.58 & 37.56 & 56.93 & 8.64 & 16.54 & 5.67 \\


D$^3$Seg~\cite{ali2026pre}
& 91.92 & \textbf{85.74} & 82.03
& 4.65 & \underline{4.63} & 2.85
& 92.96 & 90.70 & 86.78
& 4.58 & 4.04 & 2.96
& \underline{89.40} & \underline{92.85} & 91.92
& 8.30 & 5.37 & 5.45
& \textbf{89.33}$^{\ast}$ & \textbf{50.40} & 55.63
& \underline{5.44} & 14.03 & 7.06 \\


IMFuse~\cite{pipoli2025fuse}
& 91.54 & \underline{85.53} & 80.83
& \textbf{3.96}$^{\ast}$ & 4.81 & 3.23
& 92.44 & 90.36 & 84.73
& 4.54 & 3.52 & 2.83
& 89.15 & 90.74 & 89.67
& \textbf{7.41}$^{\ast}$ & 6.89 & 6.90
& \underline{88.81} & \underline{49.22} & 53.77
& \textbf{4.90} & 14.79 & 8.65 \\


mmFormer~\cite{zhang2022mmformer}
& 91.95 & 84.42 & \underline{82.59}
& 4.52 & 5.18 & 3.28
& 93.14 & 90.53 & 87.63
& 4.21 & 3.59 & 2.66
& 88.63 & 90.92 & 90.35
& 8.48 & 6.25 & 6.15
& 88.61 & 44.96 & 56.01
& 5.98 & 14.43 & 6.95 \\


SegMamba~\cite{xing2025segmamba}
& 91.74 & 81.63 & 80.74
& \underline{4.14} & 5.82 & 2.47
& \underline{93.73} & 90.04 & 87.82
& \textbf{3.49}$^{\ast}$ & 3.18 & \underline{1.76}
& 89.18 & 92.18 & \underline{92.46}
& \underline{8.25} & \underline{4.55} & \underline{4.42}
& 87.78 & 38.60 & 56.08
& 7.05 & 14.54 & 6.76 \\


SegMambaV2~\cite{xing2025segmambav2}
& 91.64 & 83.36 & 79.76
& 5.24 & 5.09 & 2.72
& 93.65 & 90.35 & 87.54
& 3.88 & \underline{3.08} & 1.91
& 89.06 & 92.36 & 92.35
& 8.85 & 4.86 & 4.66
& 87.71 & 42.26 & 55.48
& 6.66 & \underline{13.88} & 6.63 \\


UniME~\cite{song2026uni}
& \textbf{92.57}$^{\ast}$ & 84.54 & 80.04
& 5.06 & 6.60 & \underline{2.23}
& \textbf{93.97}$^{\ast}$ & \textbf{91.35}$^{\ast}$
& \underline{88.08}
& 4.78 & 3.29 & 1.82
& 88.66 & 91.80 & 91.54
& 9.68 & 5.82 & 6.14
& 87.68 & 42.33 & \underline{56.95}
& 7.61 & 14.96 & \underline{5.22} \\


\hline

\rule[-3pt]{0pt}{12pt}Proposed
& \underline{91.99} & 84.74 & \textbf{83.95}$^{\ast}$
& 4.62 & \textbf{4.48}$^{\ast}$ & \textbf{2.16}
& 93.33 & \underline{91.07} & \textbf{88.74}$^{\ast}$
& \underline{3.87} & \textbf{2.70}$^{\ast}$ & \textbf{1.65}
& \textbf{90.45}$^{\ast}$ & \textbf{93.14}$^{\ast}$
& \textbf{93.45}$^{\ast}$
& 8.38 & \textbf{4.50}$^{\ast}$ & \textbf{3.71}$^{\ast}$
& 86.32 & 43.69 & \textbf{57.65}
& 6.62 & \textbf{12.95} & \textbf{5.04}$^{\ast}$ \\

\hline

\end{tabular}%
}

\endgroup
\end{table*}

\begin{figure}
    \centering
    \includegraphics[width=0.95\columnwidth]{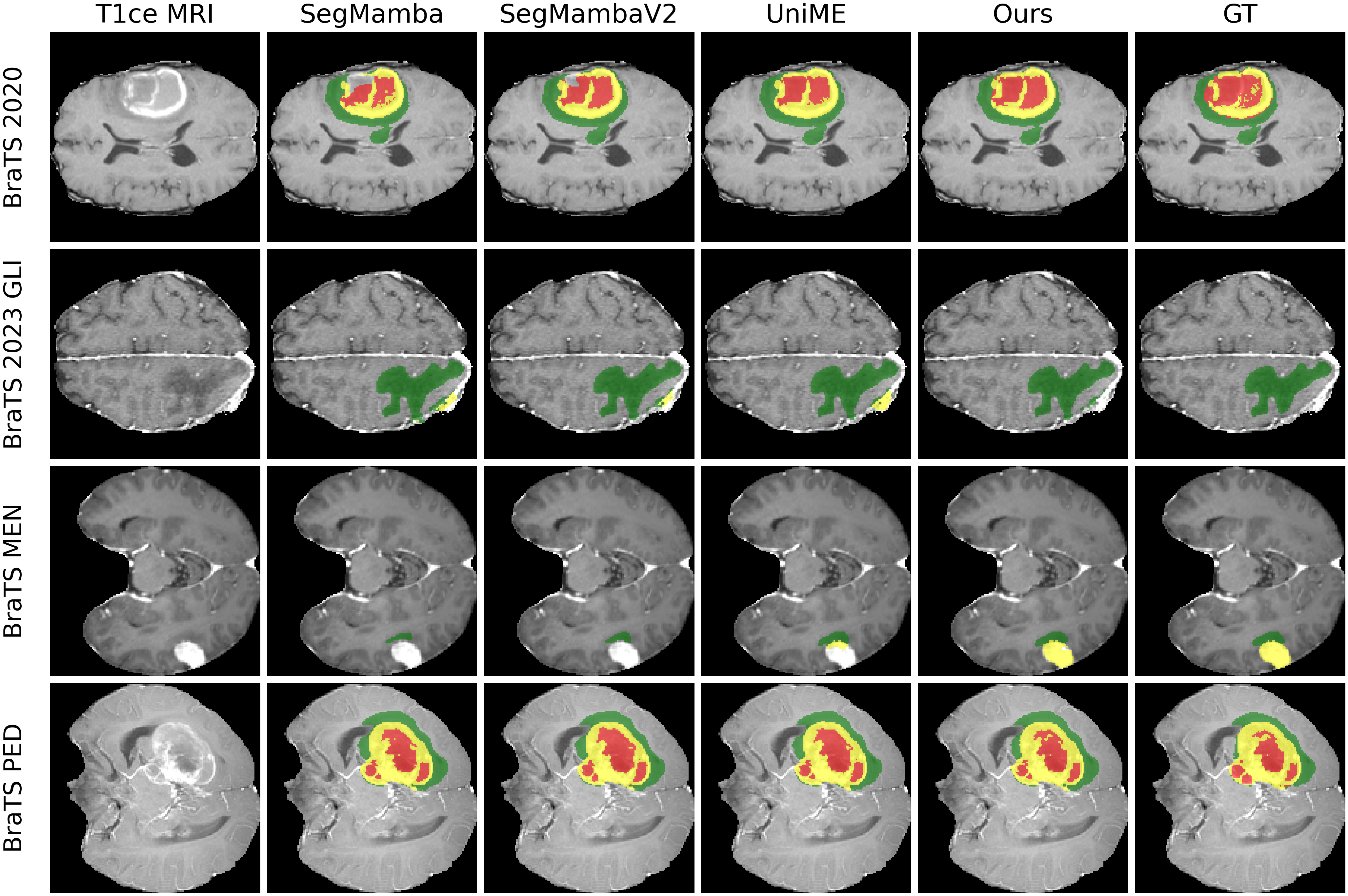}
    \caption{Qualitative comparison of segmentation predictions
across four BraTS benchmarks. Representative T1ce MRI slices
are shown alongside predictions from SegMamba, SegMambaV2,
UniME, and our proposed method, with ground-truth
annotations provided for reference. Tumor sub-regions
are highlighted as edema (green), enhancing tumor
(yellow), and necrotic/non-enhancing tumor core (red).}
    \label{fig:qualitative}
\end{figure}

\begin{table}[t]
\centering
\caption{Impact of automatically generated correction prompts
on segmentation performance across four BraTS benchmarks.
$P_1$ denotes the initial prompt-free prediction, while
$P_2$ represents the refined prediction guided by
automatically generated correction prompts without
ground-truth access at inference. Dice (\%, $\uparrow$)
and HD95 (mm, $\downarrow$) are reported for WT, TC, and ET.
Bold indicates better performance.}
\label{tab:correction_effectiveness}

\begingroup
\scriptsize
\setlength{\tabcolsep}{2.5pt}
\renewcommand{\arraystretch}{1.0}

\resizebox{\columnwidth}{!}{%
\begin{tabular}{@{}llccc|ccc@{}}
\hline
& & \multicolumn{3}{c|}{Dice (\%) $\uparrow$}
& \multicolumn{3}{c}{HD95 (mm) $\downarrow$} \\
Dataset & Method & WT & TC & ET & WT & TC & ET \\
\hline

BraTS 2020
& $P_1$ (Text-Free)
& 91.83 & 83.13 & 81.50
& 4.72 & 6.13 & 2.35 \\

Glioma
& $P_2$ (Text-Guided)
& \textbf{91.99} & \textbf{84.74} & \textbf{83.95}
& \textbf{4.62} & \textbf{4.48} & \textbf{2.16} \\
\hline

BraTS 2023
& $P_1$ (Text-Free)
& 93.30 & 90.39 & 87.82
& 4.13 & 3.35 & 1.76 \\

Glioma
& $P_2$ (Text-Guided)
& \textbf{93.33} & \textbf{91.07} & \textbf{88.74}
& \textbf{3.87} & \textbf{2.70} & \textbf{1.65} \\
\hline

BraTS 2023
& $P_1$ (Text-Free)
& 89.93 & 92.60 & 93.30
& \textbf{8.11} & 5.20 & 4.24 \\

Meningioma
& $P_2$ (Text-Guided)
& \textbf{90.45} & \textbf{93.14} & \textbf{93.45}
& 8.38 & \textbf{4.50} & \textbf{3.71} \\
\hline

BraTS 2023
& $P_1$ (Text-Free)
& 86.02 & 42.95 & \textbf{58.32}
& \textbf{5.87} & 17.25 & 5.50 \\

Pediatric
& $P_2$ (Text-Guided)
& \textbf{86.32} & \textbf{43.69} & 57.65
& 6.62 & \textbf{12.95} & \textbf{5.04} \\
\hline

\end{tabular}%
}
\endgroup
\end{table}

\section{Experiments}
\label{sec:experiments}

We use BraTS 2023 Glioma as the primary training benchmark,
comprising 1,251 multi-modal 3D MRI cases. The dataset is partitioned
into training (70\%), validation (10\%), and testing (20\%)
sets at the subject level. Each case contains four MRI
modalities (T1, T1ce, T2, and FLAIR), with annotations
for peritumoral edema (ED), necrotic/non-enhancing tumor
core (NCR), and enhancing tumor (ET). We additionally
evaluate on BraTS 2020 Glioma and two external
cohorts, BraTS 2023 Meningioma and BraTS 2023
Pediatric. The external cohorts are excluded from training, validation, and model selection to assess cross-tumor generalization without fine-tuning. For a fair comparison, all competing state-of-the-art methods listed in Table \ref{tab:tab1} are trained from scratch on the same BraTS 2023 Glioma training set, following their respective original training protocols. All methods are evaluated on identical test sets across the four BraTS benchmarks using Dice and HD95 for whole tumor (WT), tumor core (TC), and enhancing tumor (ET).

\noindent\textbf{Implementation Details.}
Our method is implemented in PyTorch using $128^3$ training patches, with data augmentation including random flipping, rotation, intensity shifting and scaling. Full-volume inference is performed using sliding-window prediction with 50\% overlap.

\noindent\textbf{Training Objectives.}
The prompt-free segmentation network (Stage 1) is trained using weighted
cross-entropy and Dice losses. The text-guided segmentation refinement network (Stage 2) is
subsequently trained using GT-derived prompts specifying
erroneous regions, directed class transitions and correction
extents, while keeping the initial network frozen.
Its principal training objectives are summarized as
\begin{equation}
\mathcal{L}_{\mathrm{corr}} =
\mathcal{L}_{\mathrm{seg}}
+\lambda_p\mathcal{L}_{\mathrm{pres}}
+\lambda_b\mathcal{L}_{\mathrm{bal}},
\label{eq:correction_loss}
\end{equation}
where $\mathcal{L}_{\mathrm{seg}}$ (combined cross-entropy and Dice losses) supervises the refined
prediction $P_2$, and $\mathcal{L}_{\mathrm{pres}}$ (KL divergence) discourages
modifications to correctly predicted regions, and
$\mathcal{L}_{\mathrm{bal}}$ (squared-deviation penalty) regularizes routing balance. The uncertainty-aware correction token generation network is then trained to predict class transitions and correction extents from
the MRI and initial prediction $P_1$:
\begin{equation}
\mathcal{L}_{\mathrm{est}} =
\mathcal{L}_{\mathrm{trans}}
+\lambda_{\mathrm{ext}}\mathcal{L}_{\mathrm{ext}},
\label{eq:estimator_loss}
\end{equation}
where the transition (class-weighted cross-entropy for transition
classification) and extent losses (Smooth L1 loss) are supervised
using GT-derived correction targets. At inference, the trained correction network automatically generates targeted correction prompts from the MRI and initial prediction
$P_1$, enabling segmentation refinement without requiring ground truth.

\begin{figure}
    \centering
    \includegraphics[width=0.95\columnwidth]{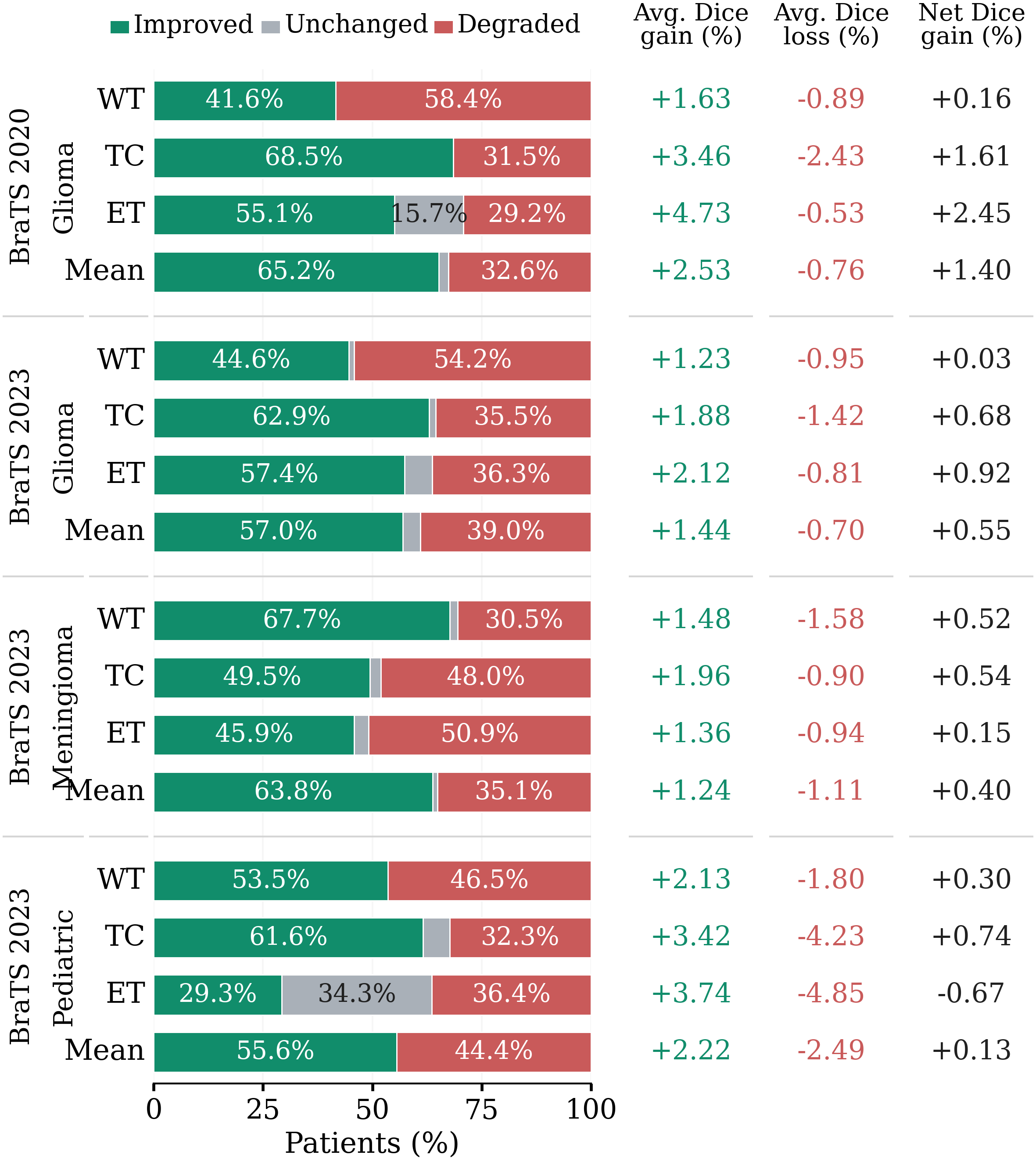}
    \caption{Patient-level analysis of automatic prompt-guided
segmentation refinement across four BraTS benchmarks.
Stacked bars show the proportions of patient cases with improved,
unchanged, or degraded Dice after refinement ($P_1$
to $P_2$). Average Dice gains and losses are computed over
improved and degraded cases, respectively, while the net gain represents the mean Dice change across all cases.}
    \label{fig:statsig}
\end{figure}

\noindent\textbf{Performance Comparison.} 
Table~\ref{tab:tab1} presents a comprehensive
comparison with seven recent state-of-the-art methods
across four BraTS benchmarks. Our method outperforms
competing approaches on 14 of the 24 evaluation metrics and ranks second best on three additional metrics.
Notably, it achieves the highest ET Dice across all four datasets. On the external Meningioma cohort,
our method achieves the highest Dice for WT (90.45\%),
TC (93.14\%), and ET (93.45\%), demonstrating promising
cross-tumor generalization.

To validate whether the observed performance gains
over competing methods are statistically significant,
we conduct paired, two-sided Wilcoxon signed-rank tests
with Holm correction ($p<0.05$). For each evaluation
metric, we compare our method against the second-best
approach when ranked first, and against the best-performing
approach otherwise. Statistically significant improvements are indicated
by an asterisk on the better-performing method in
Table~\ref{tab:tab1}. Notably, 10 of these 14 leading results are supported
by the statistical analysis, reinforcing the observed
performance gains. Conversely, competing methods
such as D$^3$Seg and IMFuse achieve higher average Dice scores
on several BraTS Pediatric metrics, but these
advantages are not consistently statistically significant. These findings underscore the importance
of statistical evaluation beyond aggregate performance
and demonstrate the superiority of our method
across diverse tumor types.

To complement the quantitative results, we present
qualitative comparisons across all four BraTS benchmarks
in Fig.~\ref{fig:qualitative}.
Compared with competing state-of-the-art methods,
our approach produces segmentations that more closely
align with the ground truth, particularly in delineating
tumor boundaries and sub-regions.
\begin{figure}[t]
    \centering
    \includegraphics[width=0.92\columnwidth]{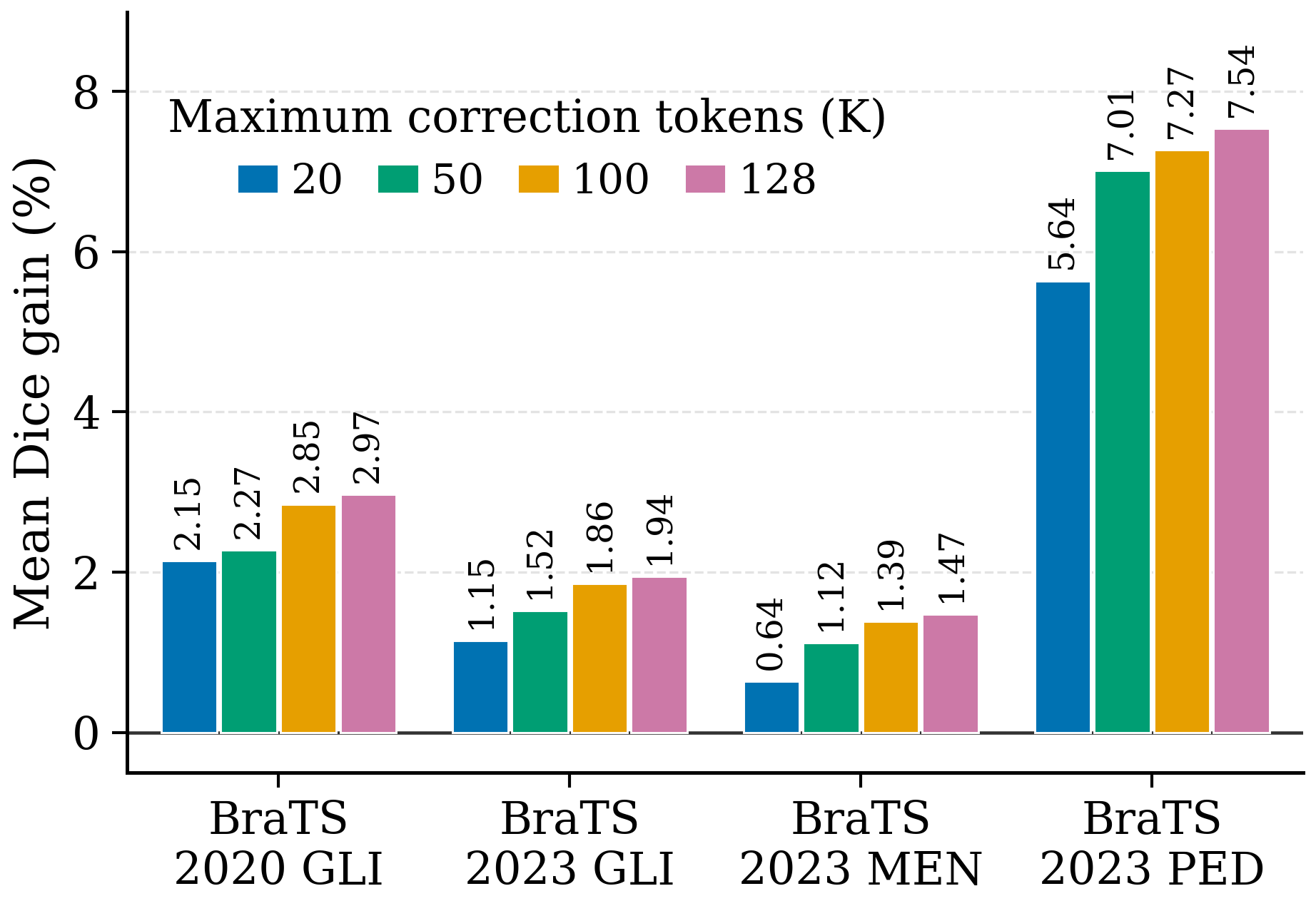}
    \caption{Impact of the number of correction tokens ($K$)
on interactive segmentation refinement across four BraTS
benchmarks. Bars represent the average/mean Dice gain (\%) across
WT, TC, and ET relative to the initial prediction $P_1$,
using GT-derived target-class instructions to simulate
clinician guidance.}
    \label{fig:ctoken}
\end{figure}

\noindent\textbf{Effectiveness of Self Correction.}
Table~\ref{tab:correction_effectiveness} evaluates the
contribution of automatic text-guided refinement by
comparing the initial prediction $P_1$ with the refined
prediction $P_2$. The refinement improves the average Dice score (computed over WT, TC, and ET) by 1.40\%, 0.55\%, 0.40\%, and 0.13\% on BraTS 2020 Glioma, BraTS 2023
Glioma, BraTS 2023 Meningioma, and BraTS 2023 Pediatric respectively.

To further examine patient-level reliability,
Fig.~\ref{fig:statsig} presents the
proportions of improved, unchanged, and degraded segmentation patient cases,
alongside their average Dice gains and losses.
Notably, refinement improves average/mean Dice in 65.2\%,
57.0\%, 63.8\%, and 55.6\% of patient cases on BraTS 2020 Glioma, BraTS 2023 Glioma, BraTS 2023 Meningioma, and BraTS Pediatric, respectively. Moreover, the average gain
among improved patient cases exceeds the average loss among
degraded cases on three of the four datasets, indicating
that the benefits of correction generally outweigh its
potential errors. An exception
is observed for ET dice on the BraTS Pediatric dataset,
where refinement results in a slight performance decrease.

\noindent\textbf{
Clinician-in-the-Loop Refinement Capability.}
We further evaluate the refinement capability of the
proposed Windowed Masked Correction Attention (WMCA)
module under varying levels of user guidance replicating clinicians-in-the-loop framework.
The model identifies uncertain spatial
cells from the initial prediction $P_1$ (see sec. \ref{interactive-sec}) and determines
their source classes from the dominant predicted labels.
Thus, user interaction is simple and limited to specifying the
desired target classes for these selected cells.

To simulate this interaction, we obtain target classes
from Ground Truth annotations. We evaluate different correction
token budgets and report the resulting improvement in
mean Dice (averaged over WT, TC, and ET) relative to
the initial prompt-free prediction $P_1$.
As shown in Fig.~\ref{fig:ctoken}, guiding refinement
with the correct target classes for only the 20 most
uncertain spatial cells selected from $P_1$ achieves
mean Dice gains of 2.15\%, 1.15\%,
0.64\%, and 5.64\% on BraTS 2020 Glioma,
BraTS 2023 Glioma, BraTS Meningioma, and BraTS Pediatric datasets
respectively. Increasing the correction token budget to 128,
corresponding to the top $K=128$ spatial cells ranked
by uncertainty in $P_1$, further raises these gains to 2.97\%, 1.94\%, 1.47\%,
and 7.54\%. 
These results demonstrate the ability of our proposed
WMCA module to interactively refine segmentations with minimal user intervention. By automatically identifying
uncertain regions and their source classes, the model
requires users (e.g., clinicians) to specify only the correct target classes, potentially reducing manual delineation effort and easy translation to medical practice.
Additional results are provided in the
supplementary material.

\noindent\textbf{Ablation Study.}
Table~\ref{tab:component_ablation} evaluates the impact of
language conditioning and correction mechanisms on
BraTS 2023 Glioma. The CD-only setting applies
language-derived keys and values (K/V) exclusively
to the correction decoder, whereas CD+CCM additionally
incorporates sparse correction token (SCT) embeddings
for bottleneck refinement without language-derived
K/V in CCM. The full model integrates both SCT
embeddings and language-derived K/V within CCM,
while retaining decoder text conditioning.

Compared with initial segmentation, CD-only configuration improves
TC and ET Dice by 0.40\% and 0.39\%,
respectively. Adding SCT-based bottleneck refinement
 further improves TC Dice by 0.15\%, with a marginal improvement of 0.06\% in ET Dice, while joint
conditioning (language + SCT) achieves the highest TC
(91.07\%) and ET (88.74\%) Dice. 
Compared with the full model, which employs four
adaptively routed experts, using only a single expert
increases WT Dice from 93.33\% to 93.41\%, but
decreases TC and ET Dice by 0.16\% and 0.45\%, respectively. Similarly, removing
TSDA from Stage 3 reduces WT, TC, and ET Dice by 0.51\%, 0.17\%, and
0.16\% respectively.

\begin{table}[t]
\centering

\caption{Ablation study evaluating the effects of
Sparse Correction Token Embeddings (SCT),
Language-derived Conditioning (Text K/V),
Stage 2 Correction Decoder (CD), Stage 2 Correction Context Module (CCM), and Stage 3 Transition-Specific Directed Attention (TSDA)
on BraTS 2023 Glioma.
Dice scores (\%) are reported for WT, TC, and ET.}

\label{tab:component_ablation}
\setlength{\tabcolsep}{2.5pt}
\renewcommand{\arraystretch}{1.0}
\resizebox{\columnwidth}{!}{%
\begin{tabular}{@{}lcccccc@{}}
\hline
\textbf{Configuration}
& \shortstack{\textbf{SCT}}
& \shortstack{\textbf{CD}\\\textbf{Text K/V}}
& \shortstack{\textbf{CCM}\\\textbf{Text K/V}}
& \textbf{WT}
& \textbf{TC}
& \textbf{ET} \\
\hline

Initial Segmentation ($P_1$)
& $\times$ & $\times$ & $\times$
& 93.30 & 90.39 & 87.82 \\

CD only
& $\times$ & $\checkmark$ & $\times$
& 93.02 & 90.79 & 88.21 \\

CD + CCM
& $\checkmark$ & $\checkmark$ & $\times$
& 93.08 & \underline{90.94} & 88.27 \\
Full Model
& $\checkmark$ & $\checkmark$ & $\checkmark$
& \underline{93.33} & \textbf{91.07}
& \textbf{88.74} \\
Full (Single Expert)
& $\checkmark$ & $\checkmark$ & $\checkmark$
& \textbf{93.41} & 90.91 & 88.29 \\

Full (w/o TSDA)
& $\checkmark$ & $\checkmark$ & $\checkmark$
& 92.82 & 90.90 & \underline{88.58} \\

\hline
\end{tabular}%
}
\end{table}

\section{Conclusion}
\label{sec:conclusion}
In this work, we presented an uncertainty-aware approach for brain tumor segmentation that improves prediction accuracy through targeted, language-guided error correction.  Our proposed four-stage network enables fully automated segmentation and refinement using automatically generated correction prompts, with an additional clinician-in-the-loop pathway for interactive correction of uncertain regions by specifying only the target classes. Extensive evaluations across four BraTS benchmarks demonstrated that our method outperforms state-of-the-art approaches in multiple Dice and HD95 comparisons, with statistical significance confirmed by paired Wilcoxon signed-rank tests. Experiments with varying numbers of correction tokens further demonstrated the effectiveness of interactive refinement, highlighting uncertainty-aware error identification and language-guided refinement as a promising direction for more accurate and reliable brain tumor segmentation.
\bibliographystyle{IEEEtran}
\bibliography{references}

\end{document}